\documentclass[twocolumn,
prl 
,superscriptaddress
]{revtex4-1}

\usepackage{graphicx}
\graphicspath{{../FiguresFinal/}}
\usepackage{subfloat}
\usepackage{subfigure}
\usepackage{amssymb}
\usepackage{amsmath}
\usepackage{color}
\definecolor{BV}{rgb}{0.1,0.,0.6}
\definecolor{R}{rgb}{0.9,0,0}
\definecolor{G}{rgb}{0.2,0.8,0.2}
\usepackage{comment}
\usepackage{multirow}

\begin{document}


\title{Quantum flutter of supersonic particles in one-dimensional quantum liquids}

\author{Charles J. M. Mathy}
\affiliation{ITAMP, Harvard-Smithsonian Center for Astrophysics, Cambridge, Massachusetts 02138, USA}
\affiliation{Department of Physics, Harvard University, Cambridge, MA 02138, USA}

\author{Mikhail B. Zvonarev}
\affiliation{Department of Physics, Harvard University, Cambridge, MA 02138, USA}
\affiliation{Univ Paris-Sud, Laboratoire LPTMS, UMR8626, Orsay, F-91405, France}
\affiliation{CNRS, Orsay, F-91405, France}

\author{Eugene Demler}
\affiliation{Department of Physics, Harvard University, Cambridge, MA 02138, USA}

\date{\today}

\begin{abstract}
The non-equilibrium dynamics of strongly correlated many-body systems exhibits some of the most puzzling phenomena and challenging problems in condensed matter physics. Here we report on essentially exact results on the time evolution of an impurity injected at a finite velocity into a one-dimensional quantum liquid. We provide the first quantitative study of the formation of the correlation hole around a particle in a strongly coupled many-body quantum system, and find that the resulting correlated state does not come to a complete stop but reaches a steady state which propagates at a finite velocity. We also uncover a novel physical phenomenon when the impurity is injected at supersonic velocities: the correlation hole undergoes long-lived coherent oscillations around the impurity, an effect we call quantum flutter. We provide a detailed understanding and an intuitive physical picture of these intriguing discoveries, and propose an experimental setup where this physics can be realized and probed directly.
\end{abstract}

\maketitle
Quantum environments are known to be capable of drastically altering the properties of embedded particles, notable examples being the formation of polarons in solid-state systems \cite{Alexandrov_book}, Kondo singlets in systems with localized impurities \cite{Hewson_book}, and quasiparticles in Fermi liquids \cite{Nozieres_book}.
The study of these phenomena has typically been carried out assuming the dressing of the particle is in equilibrium \cite{girardeau_impurity_TG_09}, however recent experiments have begun addressing nonequilibrium phenomena associated with the formation of these strongly correlated states \cite{Latta2011,Loth_2010,palzer_impurity_transport_09}. The theoretical analysis poses one of the most formidable challenges in condensed matter physics, requiring accurately capturing the dynamics of strongly interacting quantum phases of matter \cite{Caux09,Werner08,Schollwoeck11}. In this work we provide an essentially exact numerical study of the formation of a correlation hole around an impurity injected into a one-dimensional gas of hardcore bosons, also known as the Tonks-Girardeau (TG) gas \cite{tonks_36,girardeau_impurity_TG_60}. Intriguingly, we find the most striking features when the particle is injected supersonically. The physics of fast particles is responsible for a rich variety of phenomena, such as flutter in aerodynamics, Cerenkov radiation \cite{Cerenkov_book}, and bremsstrahlung\cite{Kalinovskii_book_1d}.
Our results provide an example of new physics induced by supersonic motion in a non-relativistic quantum system. Previous works on fast propagation in Bose gases assumed either a weakly coupled gas described by a set of noninteracting Bogoliubov excitations
\cite{Schecter11,Hakim96,Kamchatnov2008,Carusotto2006,astrakharchik_heavy_impurity_04,Rutherford11}, or a strongly interacting system treated within a low-energy effective field theory approach \cite{zvonarev_ferrobosons_07,imambekov_phenomenology_09,imambekov_universal_09}.
In this paper we find that there are novel features that require both the strong coupling regime and a high energy impurity, thus going beyond the regime addressed in previous works.

Our main observations in tracking the fate of the impurity injected into a 1D quantum liquid are twofold. Firstly, the injected particle forms a strongly correlated state with the quantum liquid that does not come to a full stop, instead it reaches a steady state which propagates at a reduced velocity.
We provide a study of the formation of these dissipationless propagating states, and a protocol for their generation by direct particle injection.
Secondly, if the impurity is injected at a supersonic velocity, the correlation hole around the impurity undergoes pronounced oscillations. We call this phenomenon quantum flutter in analogy with supersonic flutter in aerodynamics, which also arises from nonlinear interactions of a fast object with the background medium.
This quantum flutter is due to the formation of an entangled many-body state, whose coherence is long-lived.
Recent work has shown that strongly coupling a particle to a bath can lead to non-Markovian dynamics and the possibility of coherence surviving for long times \cite{Liu2011,Chru2010}.
Quantum flutter provides an example of a quantum system taken far out of equilibrium whose relaxation shows striking quantum coherent effects that go beyond a hydrodynamical description.
We propose a direct experimental realization of this physics in a cold atomic setting, where impurity physics in Tonks-Girardeau gases has already been realized \cite{palzer_impurity_transport_09,wicke_spin1D_10,catani_impurity_dynamics_11}. But first, we develop a detailed understanding of the intricate and intriguing physics of the formation of the dissipationless current-carrying state and subsequent quantum flutter.

\subsection{Physical system and correlation hole formation}

\begin{figure*}[htp]
\centering
\includegraphics[width=1\linewidth, clip=true,trim= 0 0 0 0]{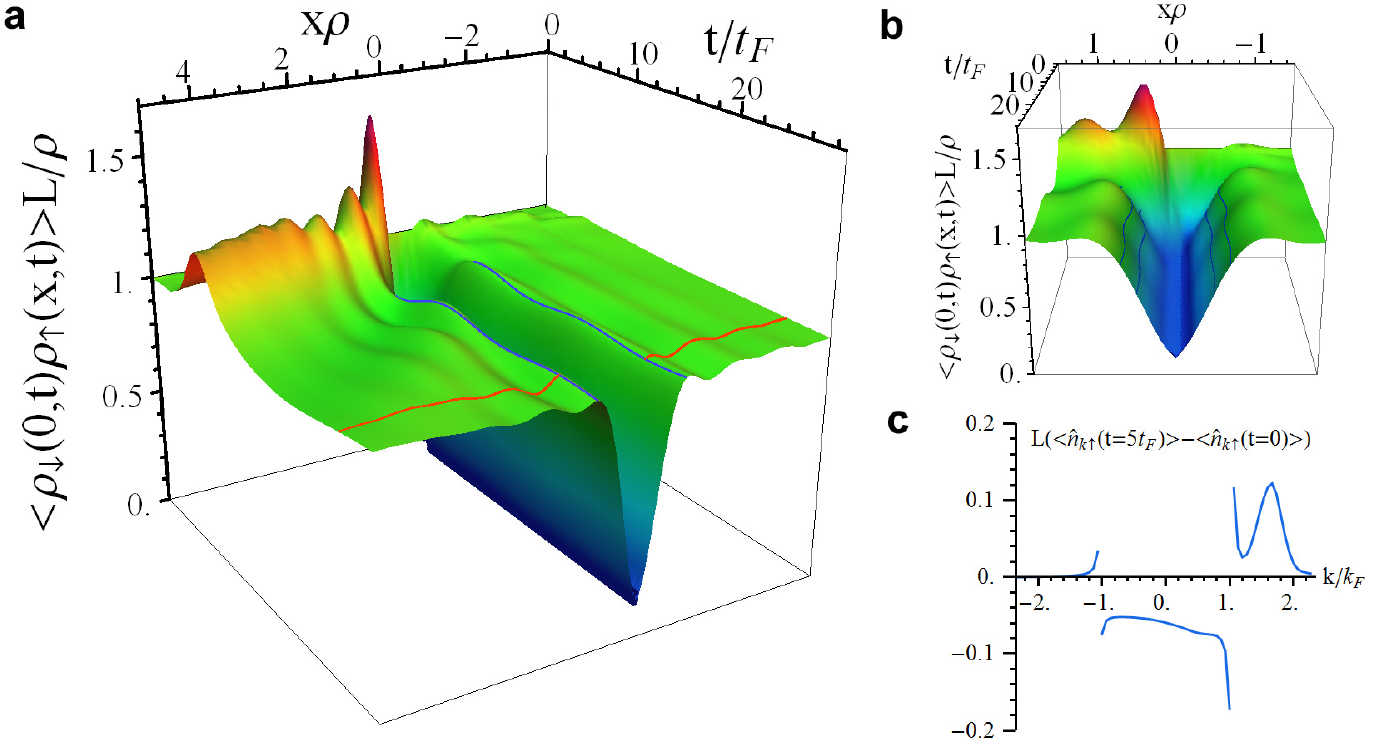}
\caption{
\textbf{Correlation hole formation and quantum flutter}:
Formation of the correlation hole and emission of the wave packet in the background gas, for interaction strength $\gamma=5$ and injected momentum $Q=1.35 k_F$.
\textbf{a},
Time evolution of the density distribution around the impurity, $L \langle \hat{\rho}_{\uparrow}(x,t) \hat{\rho}_{\downarrow}(0,t)\rangle/\rho$, for an impurity with initial momentum $Q=1.35 k_F$ pointing in the $x>0$ direction.
From $t=0$ to about $t=5 t_F$, the correlation hole forms. Simultaneously, a large wave packet in the background gas forms (red ridge). Friedel-like oscillations in the spatial direction are visible outside of the correlation hole.  We highlighted the line at $t=25t_F$ outside the correlation hole in red, where the oscillations are clearly visible.
In blue we highlighted two lines at constant positions ($x\rho=-0.85$ and $x\rho=0.93$) on the peaks bordering the correlation hole, along which oscillations, which we call quantum flutter, are visible.
\textbf{b},
Front view of the correlation hole. Here we see oscillations inside the correlation hole, which correspond to the quantum flutter. We highlighted lines of constant height ($x\rho=0.6$ and $0.8$), which are oscillating back and forth in space as a function time, thus the correlation hole is 'fluttering'.
\textbf{c},
Change in momentum distribution in the background gas after the formation of the correlation hole: \mbox{$L\Big(\langle\hat{n}_{k\uparrow}(t=5t_F)\rangle-\langle\hat{n}_{k\uparrow}(t=0)\rangle\Big)$} where $\hat{n}_{k\uparrow}=c^{\dagger}_{k\uparrow} c_{k\uparrow}$.
A narrow peak in the momentum distribution of the majority particles is formed (indicated by the red arrow), corresponding to the emitted wave packet. Also, we see a depletion of $\langle \hat{n}_{k\uparrow} \rangle$ for $|k|<k_F$, from the correlation hole being delocalized across the Fermi sea. Note that $\langle \hat{n}_{k\uparrow} \rangle$ is defined in the fermionic language. }
\label{fig:Corr}
\end{figure*}

Our system consists of an impurity with flipped spin of mass $m_{\downarrow}$ interacting via a contact interaction with a one-dimensional TG gas of hardcore bosons, of mass $m_{\uparrow}$. At present, we assume the masses are equal: $m_{\uparrow}=m_{\downarrow}=m$. In the absence of the impurity, Girardeau \cite{girardeau_impurity_TG_60} showed that the TG gas can be mapped to a fully polarized noninteracting Fermi gas, thus solving an interacting many-body quantum problem exactly.  We have now extended this argument to the problem with an impurity: because the impurity is distinguishable from the TG bosons, the TG bosons can be fermionized independently of the impurity.

Since we are describing results for an impurity in a TG gas and a fully polarized Fermi sea simultaneously, we can adopt the notation from the fermionic system. All the 'fermionic' quantities we describe have direct analogs in the TG gas. For example, we define the 'Fermi momentum' $k_F$ of the TG gas (or the Fermi sea) through its density $\rho$: $k_F=\pi\rho$. We will refer to either the TG bosons or the fullly polarized fermions as the background particles, and give them all spin up. The impurity is then a down spin.
We define fermionic creation (annihilation) operators: $c^{\dagger}_{k\sigma}$ ($c_{k\sigma}$) creates (annihilates) a fermion of momentum $k$ and spin $\sigma$ ($\sigma=\uparrow$ or $\downarrow$).
The ground state of the background gas becomes simply a fully polarized Fermi sea: $|\mathrm{FS}\rangle=\prod_{|k|<k_F} c^{\dagger}_{k\uparrow} |0\rangle$, where $\mathrm{FS}$ stands for Fermi sea and $|0\rangle$ is the vacuum.

We call $x_{\downarrow}$ the position of the impurity, and $x_1, \ldots, x_N$ the positions of the N particles in the background gas (where N is odd).
The Hamiltonian of the system is
\begin{eqnarray}
H&=&\frac{\hat{P}_{\downarrow}^2}{2 m_{\downarrow}} +\sum_{i=1}^N \frac{\hat{P}_{i\uparrow}^2}{2m_{\uparrow}} + g \int dx \hat{\rho}_{\downarrow}(x) \hat{\rho}_{\uparrow}(x)
\label{eq:Ham}
\end{eqnarray}
where $\hat{P}_{\downarrow}$ ($\hat{P}_{i\uparrow}$) is the momentum and $m_{\downarrow}$ ($m_{\uparrow}$) the mass of the impurity ($i-th$ background particle);
g is the coupling between the impurity and the bosons;  $\hat{\rho}_{\sigma}(x)=\psi^{\dagger}_{\sigma}(x) \psi_{\sigma}(x)$ is the density operator of spin $\sigma$, where $\psi^{\dagger}_{\sigma}(x)$ ($\psi_{\sigma}(x)$) creates (annihilates) a fermion of spin $\sigma$ at position x. As stated previously we set $m_{\downarrow}=m_{\uparrow}=m$ for now.
We parametrize the interaction strength between the impurity and the background particles with the dimensionless interaction parameter $\gamma=mg/\rho$.

To obtain the physics we are describing, we start the impurity in a plane wave, thus our initial state is
\begin{equation}
|\Psi\rangle=c^{\dagger}_{Q\downarrow} |\mathrm{FS}\rangle
\end{equation}
where $Q$ is the initial impurity momentum. We let this state evolve in time under the Hamiltonian in Eq. \ref{eq:Ham} and then calculate the expectation value of the quantities we are interested in.

\begin{figure*}[htp]
\centering
\includegraphics[width=1\linewidth, clip=true,trim= 0 0 0 0]{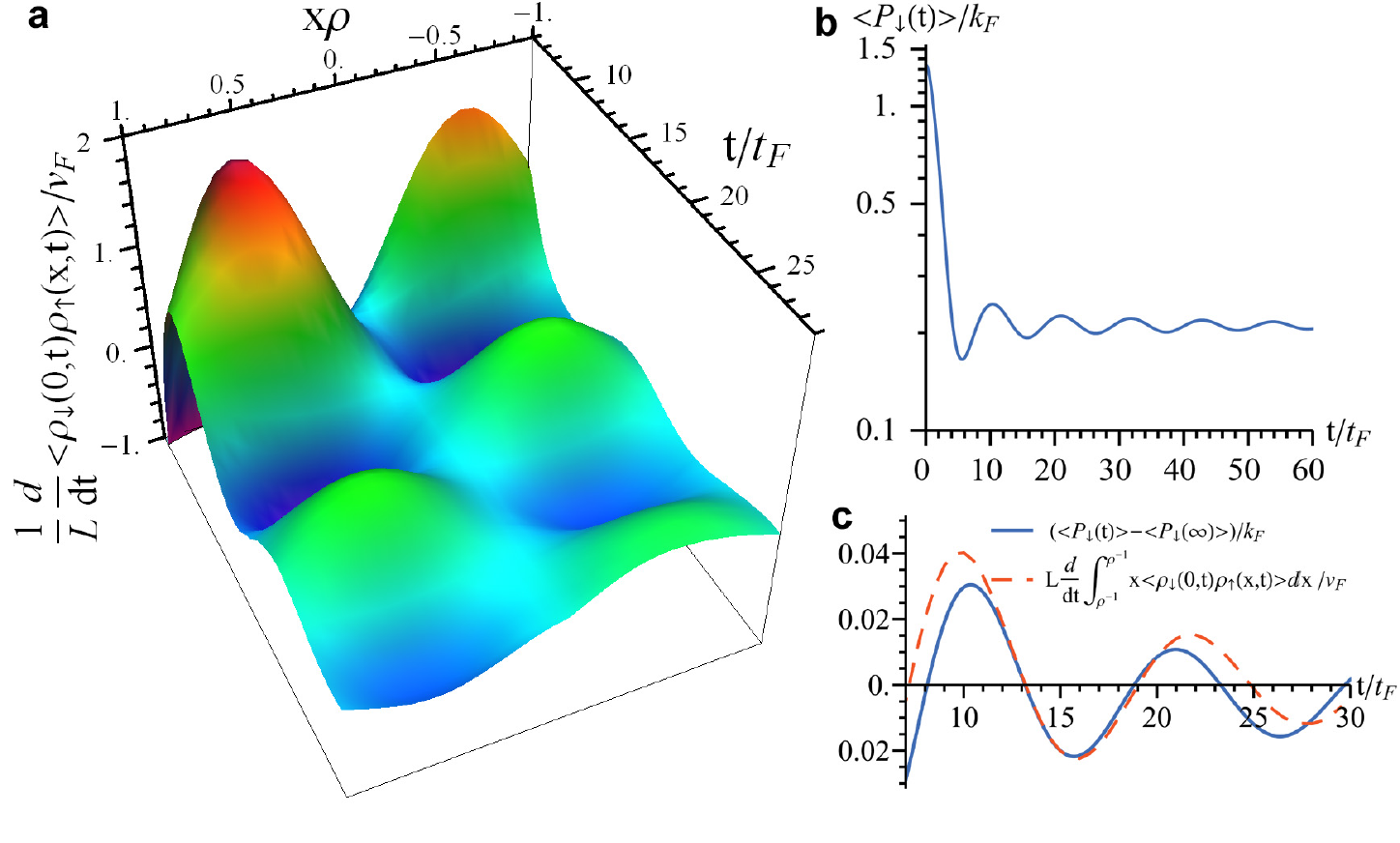}
\caption{
\textbf{Properties of quantum flutter}:
Properties of the quantum flutter, with parameters $\gamma=5$ and $Q=1.35 k_F$. We observe oscillations of the correlation hole around the impurity, and of the momentum of the impurity, which are in phase and consistent with the picture of the impurity exchanging momentum with its correlation cloud.
This suggests that coherent quantum mechanical processes are leading to a collective mode of the system, which we call quantum flutter.
\textbf{a},
Plot of the time derivative of the density distribution of the background gas within the correlation hole: $\frac{1}{L}\frac{d}{dt} \langle \hat{\rho}_{\downarrow}(0,t) \hat{\rho}_{\uparrow}(x,t)\rangle L/v_F$ with $x$ within an interparticle distance $\rho^{-1}$ of the impurity. Pronounced oscillations of the correlation hole are observed at a fixed frequency.
\textbf{b},
Time evolution of the expected momentum of the impurity. We notice two main features: first the momentum does not decay to zero. Second, the momentum shows oscillations at a fixed frequency around the saturation value. These oscillations are in phase with the oscillations of the correlation hole. 
\textbf{c},
Comparison of the time derivative of the first moment of the density of the background gas inside the correlation cloud, \mbox{$L\frac{d}{dt} \int^{\rho^{-1}}_{\rho^{-1}}dx x {\rho}_{\downarrow}(0,t) \hat{\rho}_{\uparrow}(x,t)\rangle/v_F$}, and the oscillations of the impurity momentum. If the background gas were classical, this time derivative of its first moment would correspond to its momentum. The oscillations of the two quantities are almost in phase and of the same order of magnitude, substantiating the claim that the quantum flutter corresponds to the impurity exchanging momentum with its correlation hole.
}
\label{fig:PDownsOne}
\end{figure*}

Because this system is integrable, obtaining the many-body eigenstates of the problem reduces to solving the Bethe-Ansatz equations relevant to this problem \cite{castella_mob_impurity_93,lamacraft_impurity_09}. Therefore, after obtaining all the eigenstates, one can calculate the expectation value of any operator, by introducing a complete set of eigenstates, and using massively parallelized computing resources (see Supplementary Information Sections S.1 through S.4 for details).
To probe the dynamics in our system, we calculate the density distribution of the quantum liquid around the impurity. This corresponds to calculating the expectation value
 $\frac{1}{L}\langle \hat{\rho}_{\downarrow}(0,t) \hat{\rho}_{\uparrow}(x,t)\rangle L/v_F$, which measures the density of the quantum liquid a distance x away from the impurity (we multiply by $L$ and divide by the Fermi velocity $v_F=\hbar k_F/m$ such that this quantity is 1 at time $t=0$).
We plot this quantity in Fig.~\ref{fig:Corr}a for coupling $\gamma=5$ and initial momentum $Q=1.35 k_F$. We define a Fermi time as the inverse Fermi energy: $t_F=1/E_F=2m/k_F^2$.
We see that for t from 0 to $5 t_F$, the correlation hole is forming, and a narrow wave packet in the background gas is being emitted. This last statement can be verified by calculating the momentum distribution in the background gas, $\langle \hat{n}_{k\uparrow}(t)\rangle$. We plot the change in this quantity in Fig.~\ref{fig:Corr}c at $t=5t_F$, and indeed find a narrow wave packet peaked at some momentum close to $Q$
(Note that $\langle \hat{n}_{k\uparrow} \rangle$ only corresponds to the physical momentum distribution in the fermionic case \cite{tonks_36}).

\subsection{Phenomenology of the quantum flutter}

In the pictures of the density distribution of the quantum liquid around the impurity, we saw hints of the quantum flutter as evidenced by oscillations in time inside the correlation hole (see Fig.~\ref{fig:Corr}b). To see these oscillations in more detail, we plot the time derivative of the density distribution  $\frac{1}{L}\frac{d}{dt} \langle \hat{\rho}_{\downarrow}(0,t) \hat{\rho}_{\uparrow}(x,t)\rangle L/v_F$ inside the correlation hole (meaning within an interparticle distance $1/\rho$ from the impurity). We see that the correlation hole is moving back and forth around the impurity. We can find a pronounced signature of this effect in another quantity, the time dependence of the momentum of the impurity $\langle P_{\downarrow}(t)\rangle$ which we plot in Fig.~\ref{fig:PDownsOne}b. The expected impurity momentum initially decays quickly during the formation of the correlation hole, but the decay eventually abates, leaving the impurity with a final momentum equal to a sizeable fraction of the Fermi momentum. After this initial drop, the impurity momentum shows pronounced oscillations at a fixed frequency. The first phenomenon, saturation of momentum loss, happens for any nonzero initial momentum, but the oscillations only become pronounced when $Q$ is close to or larger than $k_F$. The oscillations in the impurity momentum and the oscillations of the correlation hole are in phase, as evidenced
in the time derivative of the first moment of the density distribution of the quantum liquid around the impurity: $L \int^{\rho^{-1}}_{-\rho^{-1}} dx x \langle \hat{\rho}_{\downarrow}(0,t)\hat{\rho}_{\uparrow}(x,t)\rangle/v_F$, see Fig.~\ref{fig:PDownsOne}c. This quantity corresponds to the classical momentum of a density distribution of the form $\langle\hat{\rho}_{\downarrow}(0,t) \hat{\rho}_{\uparrow}(x,t)\rangle$, and it oscillates approximately in phase with the oscillations of the impurity momentum.
This result suggests that the oscillations are due to a momentum exchange between the impurity and its correlation hole.

\subsection{Physical mechanism behind the quantum flutter}

\begin{figure}[htp]
\centering
\includegraphics[width=1\linewidth, clip=true,trim= 0 0 0 0]{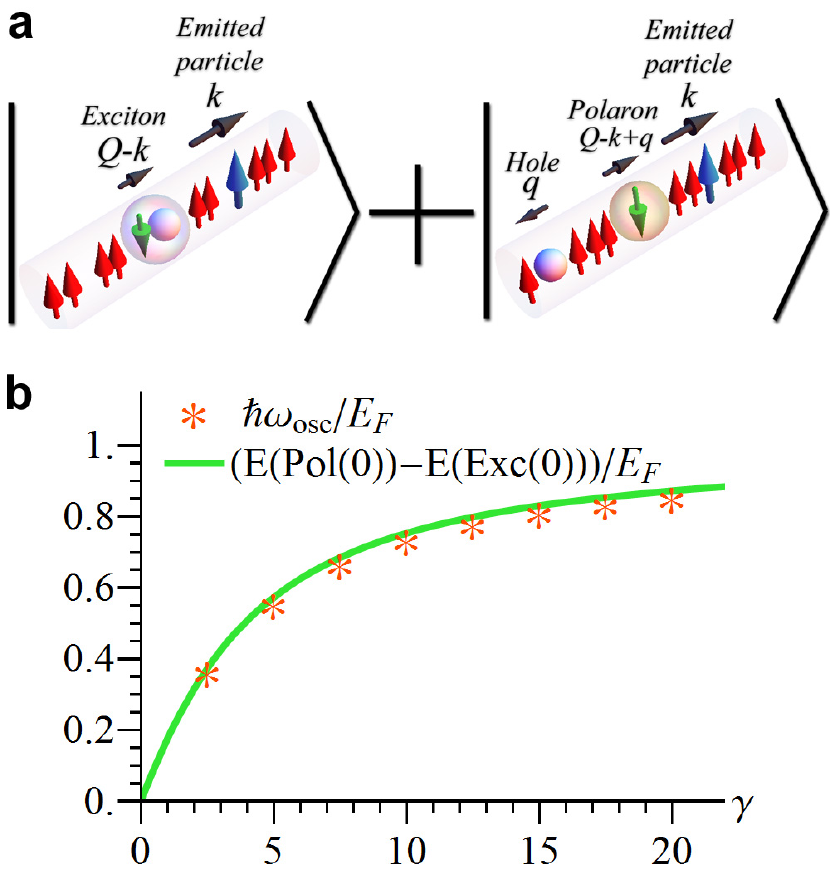}
\caption{
\textbf{Physical picture behind quantum flutter}: An intuitive picture behind the quantum flutter involves the system forming a coherent quantum superposition of two families of states. The energy difference between the van Hove singularities, where the density of states diverges, in these families of states is exactly equal to the oscillation frequency of the quantum flutter, which becomes a quantum beating effect in this picture.
\textbf{a},
The families of states behind the quantum flutter, as discussed in the text. The impurity can bind to the hole left behind by the emitted particle, creating an exciton. Alternatively, it can also not bind to the hole and formed a state dressed with particle-hole pairs, called a polaron. Therefore, the two families of states are $c^{\dagger}_{k\uparrow} |Exc(Q-k)\rangle$ (Exciton+Particle) and $c^{\dagger}_{k\uparrow} c_{q\uparrow} |Pol(Q-k+q)\rangle$ (Polaron+Hole+Particle).
\textbf{b},
Stars: oscillation frequency $\hbar w_{osc}=\hbar 2\pi / \tau_{osc}$ in the impurity momentum where $\tau_{osc}$ is the period of the oscillations seen in Fig.~\ref{fig:PDownsOne}b. Green line: energy difference between $c^{\dagger}_{Q\uparrow} |Exc(0)\rangle$ and $c^{\dagger}_{Q\uparrow} c_{0\uparrow} |Pol(0)\rangle$, which equals $E(Pol(0))-E(Exc(0))$. They agree to within $1\%$.
}
\label{fig:WaveFns}
\end{figure}

Now that we have a complete picture of the different components in the time evolution, we detail the physical mechanism underlying the quantum flutter:
the system is undergoing coherent oscillations between two families of states which we call 'exciton-like'
and 'polaron-like' (see Fig.~\ref{fig:WaveFns}a). 
Therefore the fluttering corresponds to a quantum beating between two families of states. 
That this quantum superposition is coherent for such a long time is quite remarkable, and is related to the fact that the polaron and exciton dispersions are relatively flat at strong coupling.

To provide insight into the nature of the dynamical processes and motivate the intuitive physical picture described above we supplemented the Bethe Ansatz analysis with a variational approach based on a restricted set of wave functions which have been used succesfully in previous works to study the ground state properties of impurities in cold atomic systems \cite{Chevy06,Combescot_2007,Giraud09},  and are capable of capturing the states that we conjecture are responsible for quantum fluttering. That this approach agrees quantitatively with the Bethe Ansatz results (see Supplementary Information Section S.5) strongly supports the qualitative picture we will now describe.
We emphasize, however, that all quantitative results presented in this work (except for the mass imbalance results in Fig.~\ref{fig:PDownvsrB}) were obtained from the full Bethe Ansatz equations.

Our intuitive picture behind quantum flutter relies on two types of states, an exciton and a polaron, which we describe qualitatively using variational wave functions.
The exciton state with total momentum $K$ is the lowest energy state composed of a Fermi sea, an impurity and a hole. In its simplest incarnation it has the following form:
\begin{equation}
|Exc(K)\rangle = (\sum_q \alpha^{(K)}_{q} c^{\dagger}_{K+q\downarrow} c_{q\uparrow})|\mathrm{FS}\rangle.
\label{eq:SDE}
\end{equation}
We call $E(Exc(K))$ the energy of the exciton at momentum $K$. 
 We remind the reader that Eq. \ref{eq:SDE} gives a description of the exciton in a truncated Hilbert space. 
 A full description of this state, obtained from the Bethe-Ansatz equations,
will contain an infinite number of particle-hole pair excitations. However, this level of description is enough to obtain quantitative agreement with the Bethe Ansatz calculations, as we describe later.

A polaron state is a dressed impurity state, i.e. it consists of an impurity on top of a Fermi sea, dressed by particle-hole excitations of the Fermi sea. We call $|Pol(K)\rangle$ the ground state of a system made up of a Fermi sea and an impurity with total momentum $K$, which, allowing at most one particle-hole pair excitation, is given by \cite{Chevy06} 
\begin{equation}
|Pol(K)\rangle = (\beta^{(K)} c^{\dagger}_{K\downarrow} + \sum_{kq} \gamma^{(K)}_{kq} c^{\dagger}_{K-k+q\downarrow} c^{\dagger}_{k\uparrow} c_{q\uparrow})|\mathrm{FS}\rangle.
\label{eq:Pol}
\end{equation}
We call $E(Pol(K))$ the energy of the polaron at momentum $K$.
As for the exciton, in the full Bethe Ansatz solution the polaron will contain an infinite number of particle-hole pair excitations, but capturing the polaron at this level already gives quantitatively accurate results.

Now, once the system has emitted a wave packet in the background gas, it leaves behind a hole that the impurity can interact with. Therefore the impurity and hole can form an exciton, 
and the wave function of the system would be
\begin{equation}
\textrm{Exciton+Particle state: } c^{\dagger}_{k\uparrow} |Exc(Q-k)\rangle
\end{equation}
where $k$ is the momentum of the emitted particle.
Another option is that the impurity does not bind with the hole, instead it forms a polaron with the Fermi sea. We then have a polaron, a hole and a particle in the background, giving a wave function of the form
\begin{equation}
\textrm{Polaron+Particle+Hole state : } c^{\dagger}_{k\uparrow} c_{q\uparrow} |Pol(Q-k+q)\rangle
\end{equation}
See Fig.~\ref{fig:WaveFns}a for a schematic representation of these two possibilities.

One can interpret the quantum flutter as being due to the system being in a superposition of these two possibilities: Exciton+Particle and Polaron+Particle+Hole.
The particle has momentum close to the total momentum $Q$, and the hole is a deep hole, meaning that its momentum is close to zero. Consider now the case where the particle has momentum exactly equal to $Q$, and the hole momentum exactly equal to zero. Then our two states are $c^{\dagger}_{Q\uparrow} |Exc(0)\rangle$ and $c^{\dagger}_{Q\uparrow}c_{0\uparrow}|Pol(0)\rangle$. 
Now, we can calculate the energy of $|Exc(0)\rangle$ and $|Pol(0)\rangle$ exactly using Bethe Ansatz (this means that Eqns. \ref{eq:SDE} and \ref{eq:Pol} are now dressed with an arbitrary number of particle-hole pair excitations). Using this quantity, we can look at the difference in energy between $c^{\dagger}_{Q\uparrow}c_{0\uparrow}|Pol(0)\rangle$ and $c^{\dagger}_{Q\uparrow}|Exc(0)\rangle$, which is $E(Pol(0))-E(Exc(0))$, because both states have a Fermi sea and a particle at $Q$ so those energies cancel out, and the hole at zero momentum carries no energy.
In Fig.~\ref{fig:WaveFns}b, we plot this difference in energy and compare it to the frequency of the quantum flutter $\hbar \omega_{osc}= \hbar 2\pi/\tau_{osc}$ where $\tau_{osc}$ is the period of the oscillations in impurity momentum. The two quantities match to within $1\%$, which is consistent with our physical picture of quantum flutter arising from quantum beating. 
 Note that in calculating the energy we used the fact that the emitted wave packet is spatially separated from the impurity and therefore does not interact with it. Also, we assumed the deep hole weakly interacts with the polaron, as the polaron is mostly dressed close to the Fermi points.

As we mentioned earlier, the emitted particle does not have momentum exactly equal to $Q$, and the hole on top of the polaron is not exactly at zero momentum, instead they have some spread. However, at stronger coupling the polaron and exciton have increasingly flat dispersions. 
 Therefore as long as the hole on top of the polaron is close to zero momentum, then the energy difference between $c^{\dagger}_{k\uparrow}c_{0\uparrow}|Pol(Q-k)\rangle$ (setting the hole momentum to zero for now) and $c^{\dagger}_{k\uparrow} |Exc(Q-k)\rangle$ will depend only weakly on k. Thus the emitted particle can form a wave packet without disturbing the oscillation frequency of the flutter much. However, forming a superposition of states with different energies will lead to damping of the oscillations, which is indeed what we observe. 
From this picture we can also understanding why the oscillations only appear when $Q$ is of the order of or larger than $k_F$: $Q=k_F$ is the minimum momentum necessary to be able to create a particle in the background gas (whose momentum has to be larger than $k_F$) and an exciton at zero momentum. That the oscillations become so pronounced is then due to the van Hove singularity at the top of the exciton branch (i.e. in  $E(Exc(K))$ at $K=0$).


As mentioned earlier, the picture of quantum flutter given above is qualitative, but in fact a variational approach that can capture the families of states in Eq. \ref{eq:SDE} and Eq. \ref{eq:Pol} leads to results which are in quantitative agreement with the results from Bethe Ansatz (see Supplementary Information section S.5 for details). This strongly supports the physical understanding behind quantum flutter that we expound, of a long-lived coherent superposition of exciton-like and polaron-like states.


\begin{figure}[htp]
\centering
\includegraphics[width=1\linewidth, clip=true,trim= 0 0 0 0]{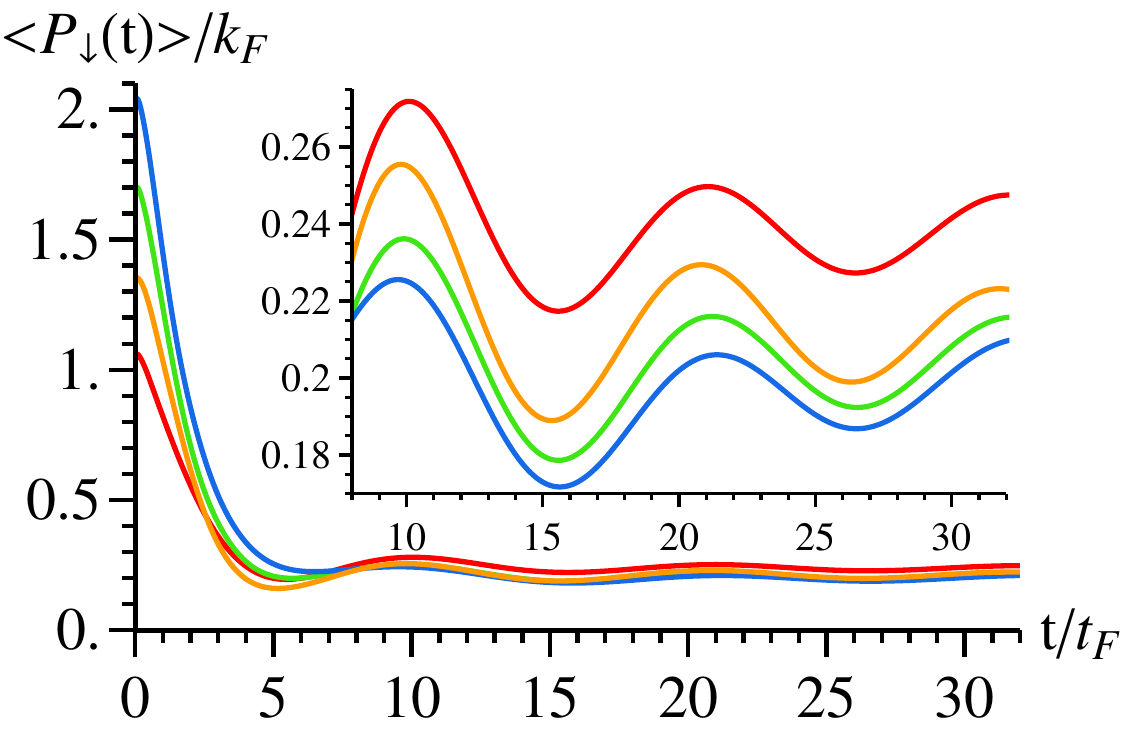}
\caption{
\textbf{Dependence of oscillations on initial momentum}:
Average impurity momentum $\langle P_{\downarrow}(t)\rangle$ as a function of time, for $\gamma=5$, for different initial momenta: (red) $Q=1.05 k_F$, (orange) $Q=1.35 k_F$, (green) $Q=1.7 k_F$, (blue) $Q=2 k_F$. Inset: zoom in on the oscillations. The structure of the oscillations depends only weakly on the initial impurity momentum. This implies (see text) that if the impurity was in a wave packet state peaked at high enough momentum, the average momentum of the impurity would flutter.}
\label{fig:PDownsvsPDown0}
\end{figure}

\subsection{Experimental consequences}

We propose several experiments to detect the physics described above.
For the most direct realization of it, consider a neutral bosonic atom that is confined to a quasi-1D geometry and made to have strong repulsive interactions via a Feshbach resonance, thus realizing a simulation of a Tonks-Girardeau gas \cite{Kinoshitab06,palzer_impurity_transport_09,palzer_impurity_transport_09,wicke_spin1D_10,catani_impurity_dynamics_11}.
This atom must have at least three internal hyperfine states $|1\rangle$, $|2\rangle$ and $|3\rangle$, such that $|1\rangle$ and $|3\rangle$ strongly interact, while the pairs $|1\rangle$ and $|2\rangle$, and the pairs $|2\rangle$ and $|3\rangle$ weakly interact. Start the system with most atoms in the $|1\rangle$ state, and a few in the $|2\rangle$ state. Now use a two-photon Raman pulse to excite the  $|2\rangle$ atoms into a $|3\rangle$ state with a given momentum. Then use time-of-flight measurements to map out the time dependence of the momentum distribution of the $|3\rangle$ particles. The averaged momentum should see the quantum flutter.


Typically impurity experiments use localized RF pulses to create impurity wave packets. From our results for $\langle P_{\downarrow}(t) \rangle$, one can immediately obtain the results for  $\langle P_{\downarrow}(t) \rangle$ for an impurity initially in a wave packet, i.e. an initial state of the form $\sum_k \alpha_k c^{\dagger}_{k\downarrow} |\mathrm{FS}\rangle$ where $\alpha_k$ is the Fourier transform of the wave packet. Namely the resulting $\langle P_{\downarrow}(t) \rangle$ can immediately be obtained from our calculations by averaging the results obtained for plane wave impurities,  weighted with $|\alpha_k|^2$:
\begin{equation}
\langle P_{\downarrow}(t) \rangle= \sum_k |\alpha_k|^2 \langle \mathrm{FS}|c_{k\downarrow} P_{\downarrow}(t) c^{\dagger}_{k\downarrow} |\mathrm{FS}\rangle.
\end{equation}
The reason is that the Hamiltonian conserves total momentum, so the different momentum components of the wave packet evolve independently of each other. 
In Fig.~\ref{fig:PDownsvsPDown0} we show that $\langle GS|c_{k\downarrow} P_{\downarrow}(t) c^{\dagger}_{k\downarrow} |GS\rangle$ is only weakly dependent on $k$ after the initial decay if $k>k_F$, therefore the quantum flutter would not be washed out if the wave packet was peaked around high enough a momentum.

\begin{figure}[htp]
\centering
\includegraphics[width=1\linewidth, clip=true,trim= 0 0 0 0]{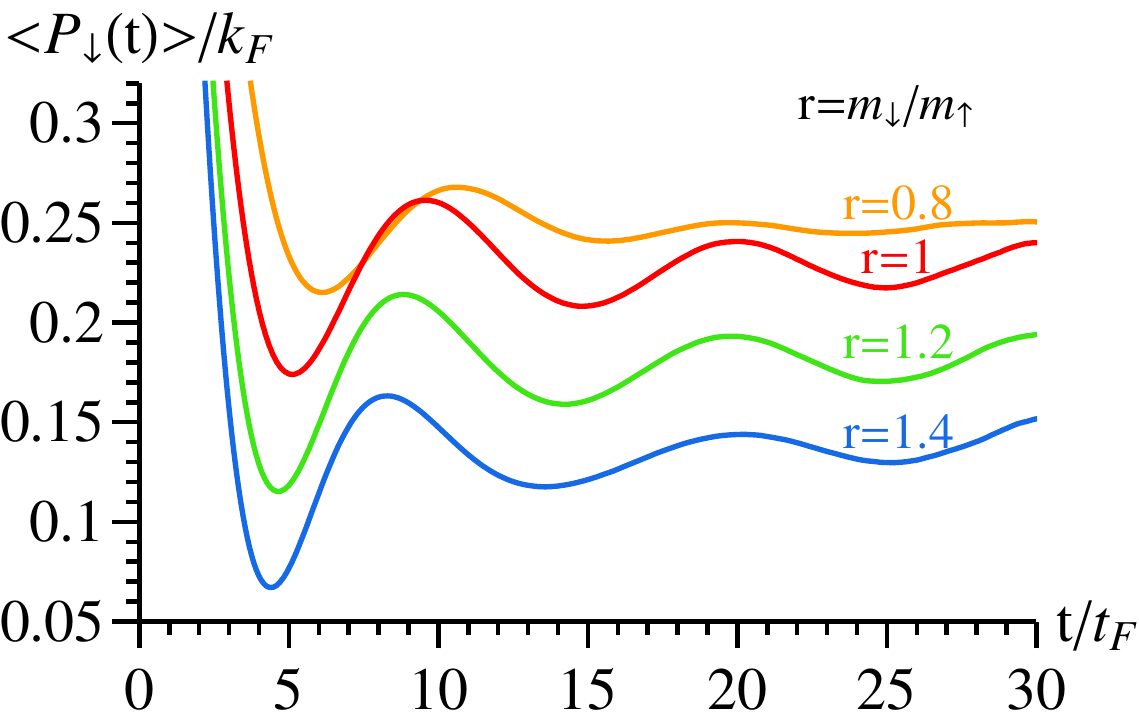}
\caption{
\textbf{Dependence of oscillations on mass imbalance}:
Time dependence of the impurity momentum $\langle P_{\downarrow}(t)\rangle$ for initial momentum $Q=1.05 k_F$, interaction strength $\gamma=5$, and different mass imbalances $r=m_{\downarrow}/m_{\uparrow}$, from top to bottom: $r=0.8,1,1.2,1.4$. These results are obtained from the variational approach discussed in the text. The $r=1$ case agrees quantitatively with the result obtained from the Bethe Ansatz approach (see Supplementary Information Section S.5). We see that the saturation of momentum loss and quantum flutter exist away from the integrability point. For $r<1$ the quantum flutter gets strongly damped, while for $r>1$ the damping depends only weakly on $r$.
}
\label{fig:PDownvsrB}
\end{figure}

Whether quantum flutter is present in other models is an important question, which we have begun answering.
As discussed above, we have a variational approach that agrees with the Bethe Ansatz results, and therefore allows us to explore non-integrable models. 
It is known that the equilibration of integrable and non-integrable models is qualitatively different, due to the infinite set of conserved quantities in the integrable case \cite{Kinoshita06}. Non-integrable models, on the other hand, are expected to relax to a local equilibrium described by hydrodynamics.
In our case, we varied the ratio $r=m_{\downarrow}/m_{\uparrow}$ of the mass of the impurity and of the background particles, making the system non-integrable. As shown in Fig.~\ref{fig:PDownvsrB} we find that the saturation of momentum loss and subsequent quantum flutter are still present when the mass ratio is changed. For $r<1$ the quantum flutter becomes more strongly damped, while for $r>1$ the damping is weakly dependent on mass ratio.

\subsection{Discussion}
We have provided an example of new physics arising from the injection of a supersonic particle into a many-body quantum system.
In many physical and biological systems \cite{Averin_book,Engel07} an important question is whether a correlated quantum state can travel through an environment while maintaining its coherence. The phenomenon of quantum flutter provides an example of the formation of an entangled state that is propagating through the many-body system and remaining coherent for long times. Furthermore, we showed that this dynamically generated and protected coherent state is not a singular feature of the integrable point, and we find similar features as we perturb away from integrability.
The most important question that remains from this work is which physical systems exhibit effects similar to quantum flutter, and more generally whether quantum systems taken far out of equilibrium exhibit quantum coherent effects, which go beyond a hydrodynamic picture of local relaxation.
We have begun answering these questions, and believe that state-of-the-art theoretical and experimental methods for studying nonequilibrium quantum dynamics should be able to shed more light on this topic.

\subsection{Acknowledgements}

\acknowledgments
We would like to thank Hans-Christof Nagerl, Elmar Haller, Michael K\"ohl, Johannes Feist, Vadim Cheianov, Dmitry Petrov, Hyungwon Kim, David Huse and Meera Parish for useful discussions.
C.J.M.M. acknowledges support from the NSF through ITAMP at Harvard University and the Smithsonian Astrophysical Observatory. M.B.Z acknowledges support from the Swiss National Science Foundation through the grant "Unconventional Regimes in One Dimensional Quantum Liquids." The numerical work was carried out on the Smithsonian High Performance Cluster.




\renewcommand{\thesection}{S.\arabic{section}}
\renewcommand{\thesubsection}{\thesection.\arabic{subsection}}

%
\makeatletter 
\def\tagform@#1{\maketag@@@{(S\ignorespaces#1\unskip\@@italiccorr)}}
\makeatother

\makeatletter
\makeatletter \renewcommand{\fnum@figure}
{\figurename~S\thefigure}
\makeatother

\renewcommand{\bibnumfmt}[1]{[S#1]}
\renewcommand{\citenumfont}[1]{\textit{S#1}}

\renewcommand{\figurename}{Figure}







\setcounter{equation}{0}
\setcounter{figure}{0}


\title{Supplementary material}




\section{Supplementary Information}


\section{The model and polaron transformation to the impurity frame \label{sect:imprefframe}}

We are considering a one-dimensional Fermi gas of atoms which we call spin up, and one impurity with spin down (in other words, we fermionized the TG gas). The second-quantized representation of the Hamiltonian is
\begin{eqnarray}
H&=& -\frac{\hbar^2}{2m} \int_0^L dx\, \sum_{\alpha=\uparrow,\downarrow} [\partial_x^2\hat\psi^\dagger_\alpha(x)]\hat\psi_\alpha(x)  \nonumber\\
&+& g \int_0^L dx\, :\hat\rho_\uparrow(x) \hat\rho_\downarrow(x): \label{eq:Hamiltonian_lab}
\end{eqnarray}
where $\hat\rho_\alpha(x) = \hat\psi^\dagger_\alpha(x) \hat\psi_\alpha(x),$ $\alpha=\uparrow,\downarrow.$ The Fourier transform
\begin{equation}
\hat\psi^\dagger_\alpha(x)= \frac1{\sqrt{L}} \sum_{q} e^{-iqx} \hat c^\dagger_{\alpha q}\quad  \label{eq:psidowninvfourier}
\end{equation}
where $\alpha=\uparrow,\downarrow$, $q= \frac{2\pi m}{L}$, $m=0,\pm1,\pm2,\ldots$,
gives the momentum-space representation of the fermion fields entering Eq.~\eqref{eq:Hamiltonian_lab}. The fermion anti-commutation relations in the momentum space are $[\hat c_{\alpha q}, \hat c^\dagger_{\beta q^\prime}]_+= \delta_{\alpha\beta}\delta_{qq^\prime}.$ The symbol ``$::$'' stands for the normal ordering.

The Hamiltonian~\eqref{eq:Hamiltonian_lab} commutes with the global spin operator $\hat{\mathbf{S}}=(\hat S_x,\hat S_y,\hat S_z)$:
\begin{equation}
[H,\hat{\mathbf{S}}]=0, \qquad \hat{\mathbf{S}}= \frac12\int_0^L dx\, \sum_{\alpha,\beta} \hat\psi_\alpha^\dagger(x) \boldsymbol\sigma_{\alpha\beta} \hat\psi_\beta(x),
\end{equation}
where $\boldsymbol\sigma =(\sigma_x,\sigma_y,\sigma_z)$ is the vector composed of the three Pauli matrices. The spin-ladder operators $S_\pm=S_x\pm iS_y$ convert spin-down particles into spin-up, and vice versa. Their momentum-space representation is
\begin{equation}
S_- = \sum_{k} \hat c^\dagger_{\downarrow k} \hat c_{\uparrow k}, \qquad S_+=S_-^\dagger. \label{eq:Spm_momentum}
\end{equation}
Another conserved quantity is the total momentum,
\begin{equation}
[H,\hat P]=0, \qquad \hat P= \hat P_\uparrow + \hat P_\downarrow. \label{eq:Hpcomm}
\end{equation}

The model~\eqref{eq:Hamiltonian_lab} is Bethe-Ansatz solvable. McGuire \cite{McGuire_Fermions1D_65} found its wave functions $|\Psi_q\rangle$ and spectrum $E_{\Psi_q}$ in the sector with one spin-down particle, and an arbitrary number $N$ of spin-up particles:
\begin{equation}
H|\Psi_q\rangle = E_{\Psi_q}|\Psi_q\rangle, \quad N_\downarrow |\Psi_q\rangle = |\Psi_q\rangle,  \quad \hat P |\Psi_q\rangle = q|\Psi_q\rangle, \label{eq:Psi_q}
\end{equation}
where $N_\downarrow= \int_0^L dx\, \hat\rho_\downarrow (x).$ Note that we write $N$ for the number of spin-up particles, which we also refer to as the host particles, instead of $N_\uparrow$ to lighten the notation. We assume $N$ is odd.

Despite the McGuire solution, the calculation of the far-from-equilibrium impurity dynamics in the model~\eqref{eq:Hamiltonian_lab} still remains a challenge. Call
\begin{equation}
|\mathrm{in}_q\rangle = c^\dagger_{\downarrow q}|\mathrm{FS}\rangle
\end{equation}
the initial state where the Fermi sea for $N$ particles is defined as ($q_m=2\pi m/L$ and 'vac' is the vacuum)
\begin{equation}
|\mathrm{FS}\rangle=\prod_{m=-(N-1)/2}^{(N+1)/2} c^{\dagger}_{q_m\uparrow}|vac\rangle.
\end{equation}
The average impurity momentum $P_q(t)$ is then
\begin{equation}
P_q(t) = \langle \mathrm{in}_q| \hat P_\downarrow(t) |\mathrm{in}_q\rangle, \label{eq:pq_labframedef}
\end{equation}
and we refer to it interchangeably as $\langle P_{\downarrow}(t)\rangle$ for shorthand.
It can be represented as
\begin{eqnarray}
P_q(t) = \sum_{\Psi_q,\Psi_{q}^\prime} &e&^{it(E_{\Psi_q} - E_{\Psi^\prime_q})} \langle\mathrm{FS}|\hat c_{\downarrow q}|\Psi_q\rangle \nonumber\\
\times &\langle& \Psi_q|\hat P_\downarrow |\Psi^\prime_q \rangle \langle \Psi^\prime_q |\hat c^\dagger_{\downarrow q}|\mathrm{FS}\rangle \label{eq:pq_labframeexpand}
\end{eqnarray}
using the completeness of the basis of the functions $|\Psi_q\rangle.$ In order to get $P_q(t)$ for sufficiently broad ranges of $q,$ $t,$ and particle number reliably one should include a vast number of terms in the sum on the right hand side of Eq.~\eqref{eq:pq_labframeexpand}. Therefore it is crucial to find a computation-efficient representation of the matrix elements entering the decomposition~\eqref{eq:pq_labframeexpand}.

To perform this task we find it convenient to use the mobile impurity reference frame. The transformation to this frame (often called polaron transformation) for an arbitrary operator $\mathcal{O}$ reads (see, for example, Ref.~\cite{castella_mob_impurity_93})
\begin{equation}
\mathcal{O} \to \mathcal{O}_\mathcal{Q} =\mathcal{Q} \mathcal{O} \mathcal{Q}^{-1}, \qquad \mathcal{Q}= e^{i\hat P_\uparrow \hat x_\downarrow} \label{eq:OtoOQ}
\end{equation}
where $\hat P_\uparrow$ is the total momentum operator of the background particles, and $\hat x_\downarrow$ is the impurity coordinate operator. This transformation acts on the impurity and host momenta as follows
\begin{equation}
\hat P_{\uparrow\mathcal{Q}} =  \hat P_\uparrow, \qquad \hat P_{\downarrow\mathcal{Q}}=  \hat P_\downarrow -  \hat P_\uparrow, \label{eq:P_imp_second}
\end{equation}
Applying the transformation \eqref{eq:OtoOQ} to the Hamiltonian of our model, we get
\begin{equation}
H_\mathcal{Q} = -\frac{\hbar^2}{2m} \int_0^L dx\, [\partial_x^2\hat\psi^\dagger_\uparrow(x)]\hat\psi_\uparrow(x) + (\hat P_\downarrow - \hat P_\uparrow)^2 + g \hat\rho_\uparrow(0). \label{eq:Hamiltonian_impurity_p}
\end{equation}
The conservation of the total momentum in the laboratory frame, Eq.~\eqref{eq:Hpcomm}, leads to the conservation of $\hat P_\downarrow$ in the impurity frame, $[H_\mathcal{Q},\hat P_\downarrow]=0.$ Therefore the functions $|\Psi_q\rangle,$ Eq.~\eqref{eq:Psi_q}, can be written as
\begin{equation}
|\Psi_q\rangle = \mathcal{Q}^{-1} c^\dagger_{\downarrow q}|0\rangle \otimes |f_q\rangle, \label{eq:fqdef}
\end{equation}
where $|0\rangle$ is the state with no particles, and the states $|f_q\rangle$ are orthogonal, normalized and form a complete basis of eigenstates of the Hamiltonian $H_\mathcal{Q}(q)$:
\begin{equation}
H_\mathcal{Q}(q) = \sum_{f_q} |f_q\rangle E_{f_q} \langle f_q |, \qquad \langle f_q|f^\prime_q\rangle = \delta_{f_q f^\prime_q}, \label{eq:orthogonality}
\end{equation}
where
\begin{equation}
H_\mathcal{Q}(q)= -\frac{\hbar^2}{2m} \int_0^L dx\, [\partial_x^2\hat\psi^\dagger_\uparrow(x)]\hat\psi_\uparrow(x) + (q - \hat P_\uparrow)^2 + g \hat\rho_\uparrow(0) \label{eq:Hamiltonian_impurity_q}
\end{equation}
is obtained by projecting the Hamiltonian~\eqref{eq:Hamiltonian_impurity_p} onto the sector with $\hat P_\downarrow$ equal to $q.$ We stress that $|\Psi_q\rangle$ and $|f_q\rangle$ are in one-to-one correspondence, and $E_{\Psi_q} =E_{f_q}.$ Substituting the expression~\eqref{eq:fqdef} into Eq.~\eqref{eq:pq_labframeexpand} we get
\begin{equation}
P_q(t) = q - \sum_{f_q,f_{q}^\prime} e^{it(E_{f_q} - E_{f^\prime_q})} \langle\mathrm{FS}|f_q\rangle \langle f_q|\hat P_\uparrow |f^\prime_q \rangle \langle f^\prime_q|\mathrm{FS}\rangle. \label{eq:pq_impframe}
\end{equation}

\section{Slater determinant representation of the wave functions}

A great advantage of the states $|f_q\rangle$ is that in the coordinate representation these states are Slater determinants:
\begin{equation}
f_q(x_1,\ldots,x_N)= \frac{Y_{f_q}}{\sqrt{N!}} \det\nolimits_N(\phi_j(x_l)). \label{eq:fqBethe}
\end{equation}
Here $Y_{f_q}$ is a normalization constant,
\begin{equation}
\phi_j(x)= \frac{1}{\sqrt L} \left[\exp\left\{i\left(\frac{2z_j}L x + \delta_j\right)\right\}-g(x) \theta_j \right] \label{eq:phiset}
\end{equation}
where $j=1,\ldots,N$ and
\begin{equation}
g(x)= \frac1\Theta \sum_{t=1}^{N+1} \exp\left\{i\left(\frac{2z_t}L x +\delta_t\right)\right\},  \label{eq:sdef}
\end{equation}
where $\Theta=\sum_{t=1}^{N+1} \theta_t$ and $\theta_t = \sqrt{a}\sin\delta_t$.
The phase shifts $\delta_t$ are
\begin{equation}
\delta_t = - \frac\pi2 + \arctan(az_t -c), \qquad t=1,\ldots,N+1 \label{eq:bd}
\end{equation}
and
\begin{equation}
a=\frac8{gL}, \qquad c=\frac{4\Lambda}{g}. \label{eq:acdef}
\end{equation}
The rapidities $z_t$ are solutions to the Bethe equations
\begin{equation}
\cot z_t= a z_t -c, \qquad t=1,\ldots,N+1 \label{eq:bethez}
\end{equation}
and
\begin{equation}
\frac{2}{L} \sum_{t=1}^{N+1} z_t =q. 
\label{eq:Pfqz}
\end{equation}
We stress that it is Eq.~\eqref{eq:Pfqz} which ensures that the total momentum of the system is $q$ and couples Eqs.~\eqref{eq:bethez} for different $t=1,\ldots,N+1.$ We will assume that $z_t$ are ordered in ascending order:
\begin{equation}
z_1\le \cdots \le z_{N+1}. \label{eq:zordering}
\end{equation}
The energy $E_{f_q}$ of the state $|f_q\rangle$ is
\begin{equation}
E_{f_q}= \frac4{L^2} \sum_{t=1}^{N+1} z_t^2. \label{eq:Efqz}
\end{equation}

Among the solutions to the Bethe equations~\eqref{eq:bethez} there is a special subset corresponding to $c=-\infty.$ One has for the solutions from this subset
\begin{equation}
z_t = \pi n_t, \quad \delta_t =0, \quad \frac{\theta_t}{\Theta}= \frac1{N+1}, \quad t=1,\ldots,N+1, \label{eq:zsing}
\end{equation}
where $n_t$ are arbitrary integers. It is thus natural to associate $z_t$ with the momenta of a free-fermion problem. Let us apply the operator $S_-$ given by Eq.~\eqref{eq:Spm_momentum} to the wave function of $N+1$ spin-up fermions:
\begin{equation}
|\Psi_q\rangle = \frac1{\sqrt{N+1}}S_- c^\dagger_{\uparrow z_1} \cdots c^\dagger_{\uparrow z_{N+1}}|0\rangle, \label{eq:Psiqsing}
\end{equation}
where $q$ is defined, as usual, by Eq.~\eqref{eq:Pfqz}. Clearly, the functions~\eqref{eq:Psiqsing} are indeed eigenfunctions of the Hamiltonian~\eqref{eq:Hamiltonian_lab}. Note also a determinant representation for $f_q$ alternative to the one given by Eq.~\eqref{eq:fqBethe}:
\begin{eqnarray}
&&f_q(x_1,\ldots,x_N)= \frac{1}{\sqrt{(N+1)!}}\nonumber\\
 &&\times\left|\begin{matrix}
\frac1{\sqrt{L}}e^{\frac{2i}L x_1 z_1} &  \cdots & \frac1{\sqrt{L}}e^{\frac{2i}L x_1 z_N} & \frac1{\sqrt{L}}e^{\frac{2i}L x_1 z_{N+1}} \\
\vdots&  \ddots & \vdots & \vdots \\
\frac1{\sqrt{L}}e^{\frac{2i}L x_N z_1}&  \cdots & \frac1{\sqrt{L}}e^{\frac{2i}L x_N z_{N}} & \frac1{\sqrt{L}}e^{\frac{2i}L x_N z_{N+1}} \\
1&  \cdots &1 &1
\end{matrix} \right|. \label{eq:fqBethesingular}
\end{eqnarray}

For $c=\infty$ we get the same set of solutions as for $c=-\infty.$ Therefore only one of the sets should be used to avoid double counting.

\section{Determinant representation of the matrix elements}

Using the identity
\begin{eqnarray}
&&\frac1{N!} \int_0^L dx_1\cdots dx_N\, \det\nolimits_N[\phi_j(x_l)] \det\nolimits_N[\varphi_j(x_l)] \nonumber\\
&=& \det\nolimits_N \left[\int_0^L dy\,\phi_j(y)\varphi_l(y) \right] \label{eq:detintid}
\end{eqnarray}
valid any functions $\phi_j$ and $\varphi_j,$ $j=1,\ldots,N$ we get the matrix elements entering Eq.~\eqref{eq:pq_impframe} is the form of determinants of $N\times N$ matrices.

To ensure the normalization $\langle f_q|f_q\rangle=1$ of the wave function~\eqref{eq:fqBethe} the constant $Y_{f_q}$ should satisfy
\begin{equation}
|Y_{f_q}|^{-2} = \frac1{\Theta^2} \left(\sum_{t=1}^{N+1} \frac{\theta_t^2}{1+\theta_t^2}\right)\prod_{t=1}^{N+1} (1+\theta_t^2), \label{eq:Ytdetresult2}
\end{equation}
where $\theta_t$ and $\Theta$ are defined by Eq.~\eqref{eq:sdef}.

The ground state wave function of $N$ free fermions, $|\mathrm{FS}_N\rangle,$ is
\begin{equation}
|\mathrm{FS}_N\rangle = \frac1{\sqrt{N!}} \det\nolimits_N \left[\frac1{\sqrt L} \exp\left\{i \frac{2u_j}L x_l \right\} \right], \label{eq:FSN>}
\end{equation}
where the momenta $u_j,$ $j=1,\ldots,N,$ are (we require $N$ to be odd)
\begin{equation}
u_j = \pi \left(-\frac{N+1}{2} + j\right), \qquad j=1,\ldots,N. \label{eq:pjFS}
\end{equation}
For the overlap of the states~\eqref{eq:fqBethe} and \eqref{eq:FSN>} the identity \eqref{eq:detintid} implies
\begin{equation}
\langle\mathrm{FS}_N|f_q\rangle =  Y_f \det\nolimits_N \mathcal{X},
\end{equation}
where the entries of an $N\times N$ matrix $\mathcal{X}$ are
\begin{equation}
\mathcal{X}^l_j = \frac{\theta_l}{\sqrt{a}} \left[\frac1{u_j-z_l} - \frac1\Theta \sum_{t=1}^{N+1} \frac{\theta_t}{u_j - z_t} \right], \qquad j,l=1,\ldots,N.
\end{equation}
The denominators on the right hand side of this equation may vanish in the case of $c=-\infty$ only. To avoid dealing with the resulting singularities it is convenient to use the representation~\eqref{eq:fqBethesingular} to calculate the overlap with the function~\eqref{eq:FSN>}. This gives
\begin{equation}
\det\nolimits_N \mathcal{X}= \frac{(-1)^{[P]}}{N+1}, \qquad  u_j= z_{P_j}
\end{equation}
and zero otherwise. Here $P$ denotes a permutation of the ordered set~\eqref{eq:zordering}:
\begin{equation}
z_1,\ldots, z_{N+1} \to z_{P_1},\ldots, z_{P_{N+1}}
\end{equation}
 and $[P]$ is the sign of the permutation.

The determinant representation of the matrix elements $\langle f_q|\hat P_\uparrow |f^\prime_q \rangle$ is
\begin{equation}
\langle f_q|\hat P_\uparrow|f^\prime_q\rangle = Y_f Y_{f^\prime} \left. \frac{\partial}{\partial \lambda} \det\nolimits_{N} (\mathcal{Y} +\lambda \mathcal{Z}) \right|_{\lambda=0}, \label{eq:fPfprime}
\end{equation}
where $N\times N$ matrices $\mathcal{Y}$ and $\mathcal{Z}$ are, respectively,
\begin{eqnarray}
\mathcal{Y}_j^l &=& \int_0^L dy\, \bar\phi_j(y)\phi_l^\prime(y) = K(z_l^\prime,z_j) - \frac{\theta_j}{\Theta} \sum_{t=1}^{N+1} K(z_l^\prime,z_t) \nonumber\\
&-&\frac{\theta_l^\prime}{\Theta^\prime} \sum_{t=1}^{N+1} K (z_t^\prime,z_j) + \frac{\theta_j \theta_l^\prime}{\Theta\Theta^\prime} \sum_{t,t^\prime=1}^{N+1} K(z^\prime_{t^\prime},z_t)
\end{eqnarray}
and

\begin{eqnarray}
\mathcal{Z}_j^l &=& -i \int_0^L dy\, \bar\phi_j(y) \partial_y\phi_l^\prime(y) \nonumber\\
&=& \frac{2}{L}  [ z_l^{\prime} K(z_l^{\prime},z_j) - \frac{\theta_j}{\Theta} \sum_{t=1}^{N+1} z_l^\prime K(z_l^\prime,z_t)
- \frac{\theta_l^\prime}{\Theta^\prime} \sum_{t=1}^{N+1} z_t^\prime K (z_t^\prime,z_j) \nonumber\\
&&+ \frac{\theta_j \theta_l^\prime}{\Theta\Theta^\prime} \sum_{t,t^\prime=1}^{N+1} z_{t^\prime}^\prime  K(z^\prime_{t^\prime},z_t) ]
\end{eqnarray}

where
\begin{equation}
K(z^\prime,z) = \frac{e^{2i(z^\prime-z)}-1}{2i(z^\prime-z)} e^{i(\delta^\prime-\delta)}, \qquad K(z,z)=1.
\end{equation}

\section{Convergence of numerics}

Now that we've derived how to calculate matrix elements like $\langle f_q | f'_q\rangle$ and $\langle f_q | f'_q\rangle$ and $\langle f_q|\hat P_\uparrow |f^\prime_q \rangle$, we discuss how to generate all the states $|f_q\rangle$. One has to generate all ordered sets of $N+1$ integers $\{n_1,\ldots,n_{N+1}\}$. We then look for a value of $c$ such that one can solve Eqns. \eqref{eq:bethez} and \eqref{eq:Pfqz}. Not all ordered sets of integers will allow for such a solution due to the quantization condition $2\sum_i z_i=qL$, therefore one should search through the space of ordered sets of integers for sets that will allow for a set or rapidities that satisfy the quantization condition.

Since the number of many-body eigenstates is exponential in the system size, the challenge is to generate eigenstates that contribute significantly to the calculation of a given quantity.
Note that although the idea is simple, one requires a clever algorithm to generate the eigenstates that contribute significantly to the calculation. Indeed, it turns out that a set of states that grows polynomially with system size is sufficient to obtain accurate results, making it tractable on massively parallelized computing resources. As was shown in other works \cite{Caux09}, we found that it is convenient to start with the set of integers symmetrically ordered around 0, and consider sets of integers obtained from this set by taking a fixed number of integers out of that set and into integers outside the set. In fact, by taking the sets of integers where either 0, 1, 2 or 3 integers have been taken out of the symmetrically ordered set, one obtains convergent results, as we show in this section. This set of of ordered integers grows polynomially with $N$, thus the complexity of the problem has become polynomial in system size and tractable.

We calculate Eq.~\eqref{eq:pq_impframe} numerically for a finite number of particles, $N$, and for a finite number of intermediate states $|f_q\rangle$ and $|f_q^\prime\rangle$ included in the sum.  In the present section we demonstrate that our choice of $N$ and of the most relevant subset of states $|f_q\rangle$ and $|f_q^\prime\rangle$ makes the results presented in the paper indistinguishable from what one would have in the thermodynamic limit (that is, for $N=\infty$) and all intermediate states included in the sum.

Let us start with discussing the dependence of $P_q(t)$ on the choice of intermediate states entering the sum in Eq.~\eqref{eq:pq_impframe}, for a given finite $N$. We define
\begin{equation}
\varrho_s = \sum_s |\langle f_q|\mathrm{FS}\rangle |^2, \quad s= \text{subset of states } \{f_q \}.
\end{equation}
The completeness of $\{f_q\}$ implies the sum rule $\varrho_{\{f_q\}}=1$ (i.e. if all states are summed over), therefore we can use the devation of $1-\rho_s$ from 0 as a check of the convergence of our numerics. Let $N_s$ be a number of states in the subset $s.$ Clearly, $\rho_s$ depends not only on $N_s$ but on which states are included in $s.$ We plot in Fig.~S1 the function $1-\rho_s$ versus $N_s$ obtained for the states ordered in a descending order of the overlap $|\langle f_q|\mathrm{FS}\rangle |.$

We then examine how $\langle P_{\downarrow}(t)\rangle$ depends on the number of states $N_s$ included in the sum in Eq.~\eqref{eq:pq_impframe}.  A typical result is illustrated in Fig.~S2. It reflects two important facts:(a) our numerics have converged;(b) the error does not increase over time, instead it remains bounded.

We can now examine the dependence of $\langle P_{\downarrow}(t)\rangle$ on $N$. The convergence of our numerics, discussed in the previous paragraphs, allows us to separate the effects of finite $N_s$ from those of finite $N.$ The latter are clearly seen in Fig.~S3. 
We see that finite $N$ effects play a role for $t$ above some critical value, which increases with $N$. At this critical time one sees revivals, which correspond to the emitted particle going around the system. Below this time the results are independent of particle number, and therefore valid for the thermodynamic limit $N=\infty.$ For example, Fig. S2b of our paper is a plot for $N=\infty.$ Indeed, the saturation of momentum loss and quantum flutter are obtained before the revivals.

\section{The variational approach}

We consider a variational wave function with two particle-hole pairs

\begin{eqnarray}
(\alpha_0(t) c^{\dagger}_Q &+& \sum_{kq} \beta_{kq}(t) c^{\dagger}_{Q-k+q} c^{\dagger}_{k\uparrow} c_{q\uparrow}\nonumber\\
&+&\sum_{kk'qq'} \gamma_{kk'qq'}(t) c^{\dagger}_{Q-k+q} c^{\dagger}_{k\uparrow} c_{q\uparrow})|\mathrm{FS}\rangle.
\end{eqnarray}
This set of wave functions can capture the states we stated were responsible for quantum flutter, Eqs. 3 and 4 in the main text.
We obtain time-dependent variational equations by minimizing the expectation value of $i \partial_t -H$ where
\begin{equation}
H=\sum_{k\sigma} \frac{k^2}{2m_{\sigma}} + \frac{g}{L} \sum_{k,k',q} c^{\dagger}_{k-q\uparrow} c^{\dagger}_{k'+q\downarrow} c_{k'\downarrow} c_{k\uparrow}.
\end{equation}
We therefore solve the time dependent equations from considering the variational minimization with respect to the variables $\alpha_0(t)$, $\beta_{kq}(t)$, and $\gamma_{kk'qq'}(t)$:
\begin{equation}
\delta \langle\Psi(t)| (i \partial_t - H)|\Psi(t)\rangle=0
\end{equation}
The resulting coupled equations can be obtained from the equations one gets from studying the ground state \cite{Giraud09}, by replacing the energy $E$ replaced with $i\partial_t$. The resulting equations contain continuum variables $k,k',q,q'$ which are discretized, and the equations are solved use Runge-Kutta integration. In Fig. S4 we compare the results obtained from the variational approach (which is in the thermodynamic limit) and the Bethe Ansatz result with 48 particles, for $\gamma=5$ and $Q=1.05 k_F$. We find that the absolute difference between the two plots remains below $0.02 k_F$.



\begin{figure*}[htp]
\centering
\includegraphics[width=0.5\linewidth]{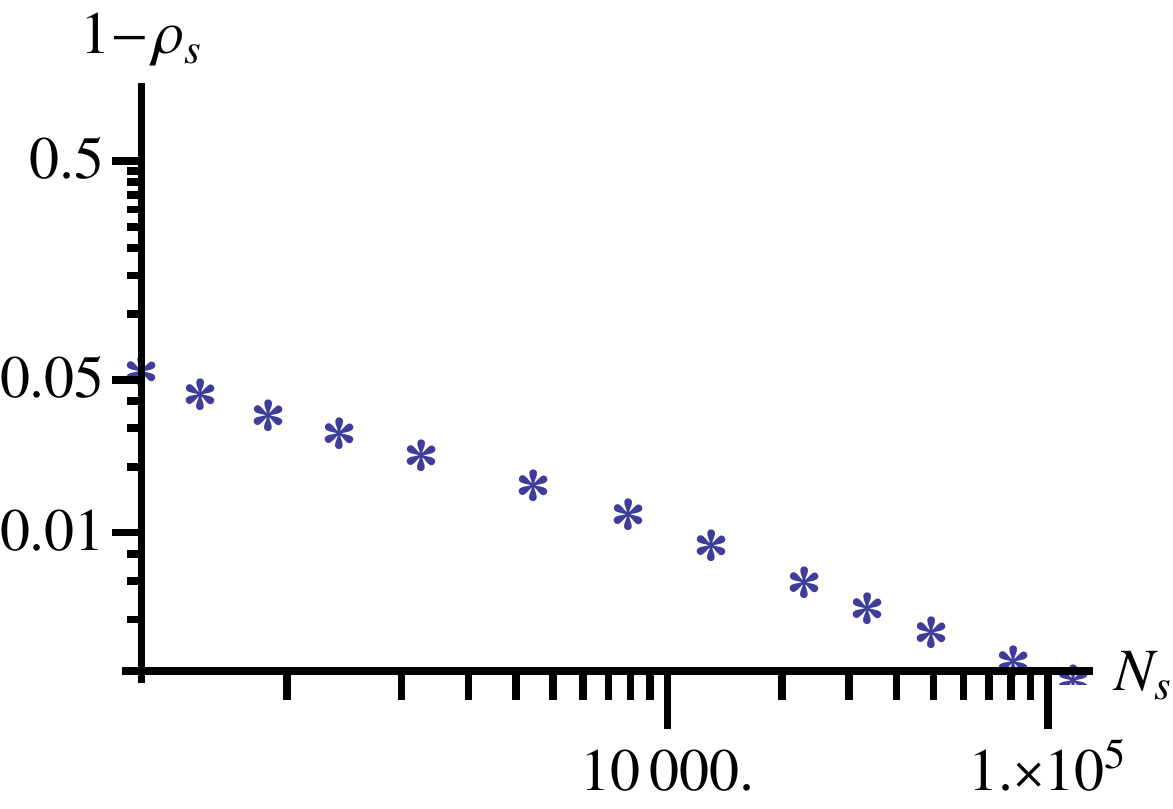}
\caption{
\textbf{Saturation of the sum rule as a function of the number of states}.
Shown is the log-log plot of the sum rule $1-\varrho_s$ versus the number of included states $N_s$. The states in the subset $s$ are ordered in decreasing order of $|\langle f_q|\mathrm{FS}\rangle |$. The parameters are: number of particles in the background gas $N=31,$ dimensionless coupling $\gamma=5,$ initial momentum $q=1.5 k_F.$ }
\label{fig:SaturationLogLog}
\end{figure*}

\begin{figure*}[htb]
\includegraphics[width=1\linewidth, clip=true,trim= 0 0 0 0]{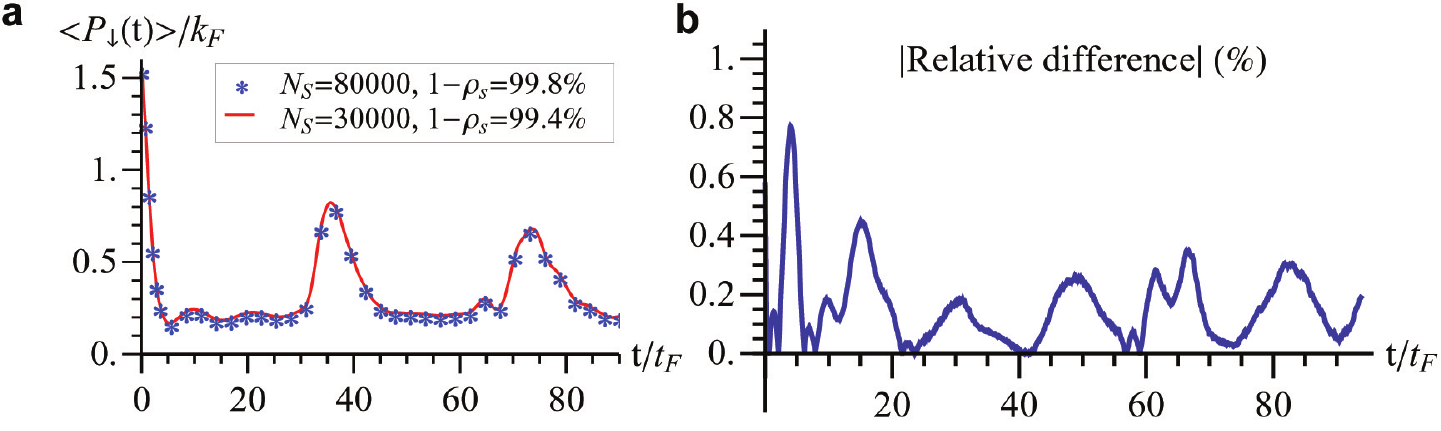}
\caption{
\textbf{Convergence of our numerical calculation of $\langle P_{\downarrow}(t)\rangle$ using the Bethe-Ansatz approach}.\\
\textbf{a},
Plot of $\langle P_{\downarrow}(t)\rangle$ for the number of included states $ N_s=3\times 10^4 (8\times 10^4)$, which corresponds to a saturation of the sum rule $\varrho_s=0.994$ (0.998).
\textbf{b}, Plot of the relative error for these two curves given on the left panel. The input parameters are the same as those used for Fig. S1: number of hosts $N=31,$ dimensionless coupling $\gamma=5,$ initial momentum $q=1.5 k_F.$}
\label{fig:CorrOsc}
\end{figure*}

\begin{figure*}[htb]
\centering
\includegraphics[width=1\linewidth, clip=true,trim= 0 0 0 0]{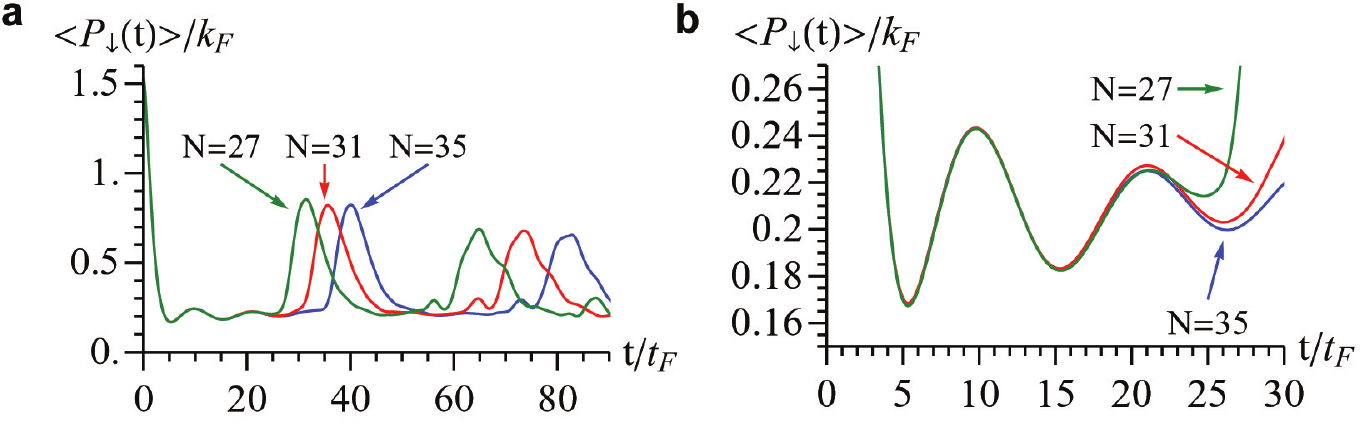}
\caption{
\textbf{Dependence of our results on the number of particle $N$ in the background gas}. We find that finite size effects manifest themselves as a revival which is pushed out to longer times for larger particle numbers. Before the revivals, the results are independent of particle number. Therefore, since the saturation of momentum loss and quantum flutter occur before the revivals, these results are valid for the thermondynamic limit $N=\infty$.
\textbf{a},
Plot of $\langle P_{\downarrow}(t)\rangle$ for $\gamma=5,$ initial momentum $q=1.5 k_F$ and various $N$: $N=28, 32, 36$.
\textbf{b},
Zoomed in plots from the left panel. We notice that the curves lie exactly on top of each other before the effects or revivals start to set in.}
\label{fig:N}
\end{figure*}

\begin{figure*}[htb]
\centering
\includegraphics[width=1\linewidth, clip=true,trim= 0 0 0 0]{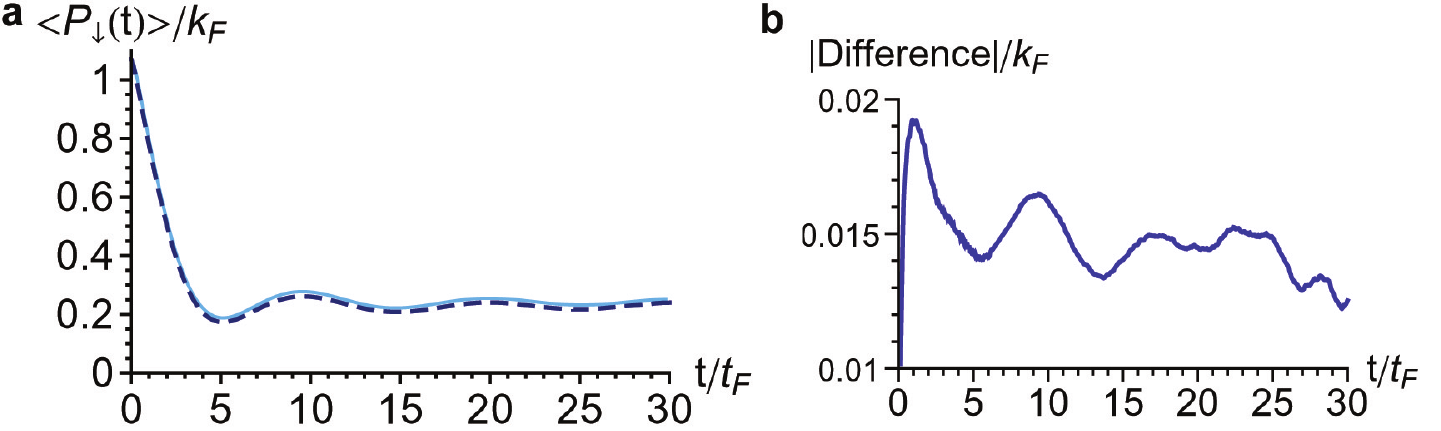}
\caption{
\textbf{Comparison of the Bethe Ansatz results with the variational approach}.
\textbf{a}, Plot of $\langle P_{\downarrow}(t)\rangle/k_F$ obtained from the Bethe Ansatz approach(full line) and from the variational approach (dotted line), for $\gamma=5$ and $Q=1.05 k_F$. The Bethe Ansatz results were obtained for 48 particles.
\textbf{b}, The absolute difference between the two curves in \textbf{a} is below $0.02 k_F$ at all times.
}
\label{fig:BAvsVar}
\end{figure*}


\end{document}